\def\LUM{\:{\rm ergs\:s^{-1}}}

\def\OIGS{\:{\rm ergs\:cm^{-2}\:s^{-1}\:\AA^{-1}}}
\def\LA{Lyman\thinspace$\alpha$}

\def\NvL{\ion{N}{5} $\lambda\lambda1239,1243$}

\def\CiiLxiii{\ion{C}{2} $\lambda1334$}

\def\CivL{\ion{C}{4} $\lambda\lambda1548,1551$}

\def\SIivL{\ion{Si}{4} $\lambda\lambda1394,1403$}

\documentclass[12pt,preprint]{aastex}

\begin{document}


\newcommand{\MSOL}{\mbox{$\:M_{\sun}$}}
\newcommand{\EXPN}[2]{\mbox{$#1\times 10^{#2}$}}
\newcommand{\EXPU}[3]{\mbox{\rm $#1 \times 10^{#2} \rm\:#3$}}  


\title{
HST Spectroscopy of the Nucleus of M33\footnote{Based on observations made with the NASA/ESA Hubble
Space Telescope, obtained at the Space Telescope Science
Institute, which is operated by the Association of Universities
for Research in Astronomy, Inc., under NASA contract NAS 5-26555.
These observations are associated with proposal \# 8341.} }

\author{Knox S. Long}
\affil{Space Telescope Science Institute, \\ 3700 San Martin Drive,
Baltimore, MD 21218}

\author{Philip A. Charles}
\affil{Department of Physics \& Astronomy,\\
 University of Southampton, \\
 Southampton SO17 1BJ,
 UK}

\and

\author{Guillaume Dubus}
\affil{California Institute of Technology, Mail Code 130-33
\\ Pasadena, CA 91125 }

\begin{abstract}

We have used Hubble Space Telescope to obtain moderate resolution
1150-5700 \AA\ spectroscopy of the nucleus of M33 and a blue star
$\sim$ 1 arcsec NNW of the nucleus in an attempt to find the
optical counterpart of the nuclear X-ray source and to
characterize stellar populations in the nuclear region.  The
STIS spectra of the nucleus can be modelled in terms of two
starbursts, one with a mass of about 9000 \MSOL\ at 40 Myrs and
the other with a mass of about 76,000 \MSOL\ at 1 Gyrs.  The blue
star is a late type O giant, with no obvious spectral anomalies to
indicate that it is associated with the luminous X-ray source. The
nuclear region is not heavily reddened; 2200 \AA\ absorption
features in the spectra of both the nucleus and the star are weak.
The data and the star formation history support the hypothesis
that the M33 nuclear source, the brightest persistent source in
the Local group, is an $\sim$10 \MSOL\ black hole binary.

Subject Headings: galaxies: individual (M33)—galaxies:
nuclei—Local Group—ultraviolet: stars galaxies : stellar content

\end{abstract}

\section{Introduction} \label{sec_intro}

The nucleus of M33 houses the most luminous persistent X-ray
source in the Local Group \citep{long1981, markert1983,long1996}.
The nuclear X-ray source is about 8 times brighter than any other
X-ray source in M33, and significantly brighter than any source in
the nuclear region of M31 \citep{supper2001}. That the brightest
source in the Local Group would be located in the nucleus of M33
was, and remains, surprising because there is no evidence that M33
contains a large central mass concentration
\citep{kormendy1993,merritt2001,gebhardt2001}.

Since visible light from the nucleus of M33 is most likely
dominated by late type stars, searches for the optical counterpart
to the X-ray source seem most likely to succeed at UV wavelengths.
To test this idea, \cite[][hereafter DLC]{dubus1999} imaged the
nuclear region of M33 with HST. These observations showed that in
the UV, the nucleus is both brighter than expected from its
optical color temperature and more compact than at long
wavelengths, as might be expected if one were observing the
accretion disk of a compact binary X-ray source. About half of the
total UV luminosity of \EXPU{3}{38}{\LUM} comes from the inner
0.14 arcsec. However, because the angular size of the nucleus is
relatively close to the limiting angular resolution of HST, it was
not possible to determine whether the emission observed at the
center is from a true point source or simply a more concentrated,
bluer stellar population \citep[see
also][]{lauer1998,matthews1999}.

\cite{oconnell1983} made one of the first attempts to understand
the stellar population history of the M33 nucleus partly in
response to the discovery of the nuclear X-ray source.  While the
appearance and mass of the nucleus resembles a globular cluster in
some ways, he concluded that about 50\% of the visible light from
the nucleus arises from stars of $<$1 Gyr in age, and that the
youngest stars are likely to have been formed about 50 Myrs ago.
Most subsequent studies have added credence to this suggestion.
For example, \cite{schmidt1990} found that the visible UV portion
of the spectrum is dominated by a near solar abundance population
of stars aged less than 500 Myrs, while at longer wavelengths an
intermediate and then older populations become progressively more
important. More recently \cite{gordon1999}, modelling both broad
band photometry from FUV to NIR wavelengths, as well as optical
and NIR spectra of M33, have suggested that nearly all of the
light arises from a burst of star formation 70 Myrs ago
``enshrouded by a shell of MW-style dust with $\tau_{\nu}\sim2$".

In an attempt to identify the optical counterpart to the bright
nuclear source in M33 and also to better characterize the source
populations in the nucleus of M33, we have now carried out
spectroscopic observations of the nucleus using the Space
Telescope Imaging Spectrograph on HST \citep{woodgate1998}
covering the wavelength range 1150 - 5700 \AA. The remainder of
this paper describes the results of this investigation.

\section{Observations} \label{sec_obs}

The STIS observations of the nucleus of M33 are summarized in
Table \ref{tab_obs}.  As indicated, the spectra were acquired in
two ``visits", one lasting 4 orbits in 1999 October, covering the
FUV (1150-1730 \AA), and the other a year later lasting 2 orbits,
covering the NUV (1570-3180 \AA) and ``visible'' (2900-5700 \AA).
These gratings were selected to cover the full wavelength range
with moderate resolving power (500-1400 depending on wavelength),
sufficient to resolve broad lines that might be expected if the
nucleus housed a mini-AGN or stars with strong P Cygni profiles
and to perform stellar population synthesis.  All of the
observations were carried out with the 0.1" x 52" slit in order to
isolate as much as possible the point source in the nucleus. The
orientation of the slit was carefully set so that spectra were
obtained of not only the nucleus, but also a 19th magnitude star
NNW of the nucleus. This was the only source, other than the
nucleus, detected within a few arcsec of the nucleus in the 1600
\AA\ (F160BW) image. (See Figure 1 of DLC.) The acquisition images
indicate that the slit was accurately centered on the nucleus for
both observations. From 2 to 8 spectra were obtained at each
grating setting; none shows evidence of any temporal variability.
In short, all of the observations appear to have been nominal.

All of the observations described here were processed using the
standard STIS pipeline and calibration files available in 2000
December.  Spectra of the nucleus and the star were extracted from
the individual exposures and combined to produce overall spectra
shown in Figures \ref{fig_nuc_data} and \ref{fig_star_data}. The
spectra from the different grating settings appear well-matched
where they overlap.

The $F_{\lambda}$ spectrum of the nucleus peaks in the UV, and
shows a prominent \LA\ absorption profile and narrow absorption
lines from a variety of ions including C II, III, and IV, Si II,
III, and IV, and Mg II.  The Balmer jump is well defined, as had
been expected from ground-based observations.  But there is no
evidence either of wind-features (P Cygni profiles) in the
spectra, nor is there evidence of obvious dust absorption in the
form of a 2200 \AA\ absorption dip. The spectrum contradicts the
suggestion by \cite{gordon1999}, based on assembling a broad-band
spectrum from UIT and HST imaging data, that a strong 2200 \AA\
feature is evident in the light of the nucleus.

The spectrum of the star is much bluer than that of the nucleus.
The low ionization state lines in the spectrum, Si II, \CiiLxiii,
and Mg II have similar strengths and widths to those in the
nuclear spectrum, suggesting they are predominantly interstellar
in both spectra. On the other hand, the highest ionization state
lines, \NvL, \SIivL, and \CivL, all show P Cygni profiles,
indicating that the star has a massive wind.

In the spectra shown in  Figures \ref{fig_nuc_data} and
\ref{fig_star_data}, the standard extraction widths were used
(0.26 arcsec for the FUV and NUV spectra and 0.35 arcsec for the
``visible'' spectra). An examination of the profiles of the
nucleus and the star along the slit shows that the nucleus is
extended compared to the star, as was anticipated. (Unfortunately
the nucleus is not sufficiently extended to easily extract
multiple spectra of the nucleus along the slit.) The flux from the
nucleus observed through the 0.1" slit is considerably less than
the flux estimated for the entire nucleus from the broad band
filters on WFPC2 \citep[see, e.g.][DLC]{gordon1999}. The ratio of
flux varies from ~1/3 at 1600 \AA\ to 1/5 at 5500 \AA. In
contrast, the STIS fluxes for the star agree within 25\% of the
broad band fluxes reported by DLC, and there is no obvious trend
with increasing wavelength.  While it is worthwhile to ask whether
this is evidence for variability in the nucleus, we view this as
very unlikely. In fact, the fluxes that we observe with STIS are
fully consistent with the results of DLC and other investigators
showing that $\sim$50\% of the UV flux arises within 0.14" of the
center of the nucleus and that the size grows at longer
wavelengths.  We have also extracted spectra of the nucleus using
extraction boxes which are somewhat larger than those provided by
the STSDAS pipeline.  The flux is somewhat (10\%) larger at all
wavelengths. Given that the nucleus is somewhat less concentrated
at longer wavelengths, one might have expected some change in the
spectral shape using larger apertures. However, there is little,
if any, change observed, and therefore we have used the standard
extractions in the analysis which follows.

\section{Analysis} \label{analysis}

The spectra of both the nucleus and the star lack dramatic
features in the UV to make identification of the source of X-rays
from the nuclear region obvious.  However, to understand whether
there are more subtle features that signify the optical
counterpart to the X-ray emission, a more detailed analysis is
required.

\subsection{Stellar Synthesis of the Nucleus}

Our STIS spectra comprise the first high quality UV spectra of the
nucleus of M33. In an attempt to model the spectra we have made
use of models generated with Starburst99 \citep{leitherer1999}. In
this initial reconnaissance, we have concentrated on the overall
shape of the spectrum.  We have primarily considered models
consisting of one or two bursts of star formation, although we
also looked at models of continuous starbursts. We have considered
models with a Salpeter initial mass function and upper mass
cutoffs 100 \MSOL\ and 30 \MSOL. Within the context of
Starburst99, we have considered models in which the metallicities
used to evolve the stellar populations range from 0.001 to 0.08.
Comparisons of the models to the data were made using $\chi^{2}$
minimization routines. Our approach was to begin with
pre-calculated models that are available from the Starview99 web
site and to use the results as guidance for calculating new models
with the version of the software and associated data available in
the spring of 2001. Starburst99 produces two sets of spectra, one
with relatively modest resolution of $\sim$10 \AA\ covering a wide
wavelength range, and one with a resolution of 0.75 \AA\ covering
the range 1205-1850 \AA. The first set of spectra are generated
with abundances that are based on model atmospheres and are
matched to that of the stellar evolution synthesis; the second are
based on the observed spectra of stars and are available only for
solar and LMC-like abundances.  Except where noted we have used
the coarser spectra in the discussion which follows.  Because the
spectra have relatively high S/N, we did not anticipate that
$\chi^{2}$ minimization would allow a strict statistical
comparison of the data or evaluation of the errors associated with
any particular model. However, it does provide a means for finding
the best parameters for a particular model, and a qualitative
measure of the degree to which the model approximates the data.


We began with a consideration of single bursts or continuous star
formation, and attempted to fit all the data from 1200-5600 \AA\
using the pre-calculated models, the coarser wavelength model
spectra, and a Galactic extinction law, specifically that of
\cite{seaton1979}. However this combination never produces
qualitatively good fits to the data.  The ``best fit'' was a
low-metallicity (z=0.004) instantaneous starburst model with an
age of 163 Myrs and an  E(B-V) of 0.23. In model fitting of this
type, the procedure selects parameters, an age, a normalization,
and the reddening, that match the continuum, including in this
case the Balmer jump. Indeed, the model approximates the spectrum
longward of 3000 \AA\ fairly well, and also produces approximately
the correct flux in the FUV region. The narrow lines, including
the Balmer lines, are not reproduced, but this is directly
attributable to the spectral resolution of the models. As shown in
Figure 3, the basic problem with this (and all other simple models
using a Galactic absorption law) is the absence, or near absence,
of a 2200 \AA\ absorption feature in the data. It is primarily
this feature that results in a $\chi^{2}_{\nu}$ of 17.8.

Galactic absorption along the line of sight to M33 is thought to
be relatively small \cite[E(B-V)$\sim$0.07;][]{vandenbergh1991},
and there is no guarantee, or possibly even expectation, that the
absorption law in M33 has a 2200 \AA\ feature as prominent as that
represented by a Seaton law with E(B-V) of 0.23. Differences in
the absorption law arise both from differences in the grain size
distribution, and in extended sources from scattering and multiple
optical depths within the source itself \citep{calzetti1994}.
Therefore we have considered two other absorption laws, neither of
which exhibits a significant 2200 \AA\ bump.  The first of these
extinction laws is that proposed by \cite{calzetti1994}. This
extinction curve was developed to account for the fact that the
2200 \AA\ feature is missing from the UV spectra of many starburst
galaxies, even though there is ample evidence for absorption from
emission line ratios. The second is for the Small Magellanic
Cloud, long known to have a smooth, but highly absorptive UV
extinction curve.  Here we use the average parameters derived by
\cite{gordon1998} for the bar of the SMC .

Allowing for all three extinction laws, the best-fit single
component model was the same z=0.004 instantaneous starburst
model, with a very similar age 143 Myrs, and a Calzetti, or
starburst, absorption law with  E(B-V) of 0.22.  The fit is
insensitive to the mass cutoff of 30 or 100 \MSOL, but models
based on evolutionary tracks for higher metallicity stars result
in higher $\chi^{2}_{\nu}$.  An instantaneous burst is better than
any of the simple continuous star formation models we tried. As
shown in Figure~\ref{fig_nuc_1}, the model fit is greatly improved
with a good match to the spectrum longward of 2300 \AA, and
relatively good approximation to the broad band FUV continuum. The
value of $\chi^{2}_{\nu}$ improves to 13.4, reflecting the higher
quality of the overall fit. The main problem qualitatively is the
structure of the spectrum shortward of 2000 \AA, which is smoother
in the model than in the observed spectrum.  Simple fits made with
an SMC-like absorption curve were generally poorer than those made
with a starburst absorption law, often worse in fact than those
made with a Galactic curve.  The underlying reason for this
appears to be associated with the fact that the UV absorption is
so strong that very low values of E(B-V) result from the model
fits, but this results in a spectral slope in the NUV and visible
that does not match the data.

The quality of the fit can be improved significantly by allowing
for two distinct bursts. The best fit model in this case is a
model synthesized from solar abundance (z=0.02) tracks and a
starburst absorption law with E(B-V) of 0.06. A relatively recent
starburst aged 70 Myrs dominates the UV light and competes with an
older burst aged 1.1 Gyrs at visible wavelengths. The improvement
in the fit to $\chi^{2}_{\nu}$ of 11.9 reflects primarily a better
approximation of the model to the FUV portion of the spectrum. The
reason that the spectrum is better approximated in the FUV
compared to the single starburst model can be traced directly to
the spectra of the higher mass stars in a younger starburst. Two
component models using either a Galactic ($\chi^{2}_{\nu}=13.8$)
or an SMC ($\chi^{2}_{\nu}=12.3$) curve yield similar parameters.
But a 2200 \AA\ problem is evident in the fits using a Galactic
absorption curve.

We have also experimented with models with three separate
components. None of these models improved the fits significantly
in a qualitative or quantitative way and all indicated that the
youngest stellar component had an age of 40-60 Myrs.

Since the two-component models described above seem to favor
models with abundances that are close to solar, we then considered
models in which the higher resolution UV stellar spectra were
substituted for the coarser model-atmospheres generated spectra in
the UV. The best fit two component model has a starburst
absorption law with E(B-V) of 0.06, a recent starburst of 50 Myrs
and an older one of 1 Gyrs. The value of $\chi^{2}_{\nu}$ drops to
9.1. Qualitatively, the fit is quite good as indicated in Figure
\ref{fig_nuc_2}.  The improvement in $\chi^{2}_{\nu}$ is clearly
due to the incorporation of higher resolution, observation-based
data in the UV range. Most of the improvement in $\chi^{2}_{\nu}$
is due to spectral resolution, although some appears due to
differences between the model-based and observation-based stellar
spectra that were used to synthesize the nuclear spectrum models.
(In fact, we have conducted experiments in which we have resampled
the data on the wavelength grid of the models, and then attempted
to fit the original data using the resampled data as the
``model''. The $\chi^{2}_{\nu}$ that results is of about 8.)

One of the remaining systematic departures now appears at between
2000 and 2400 \AA, where the model and data are discrepant at a
level of about 10\% of the continuum. This can be eliminated if we
allow for Galactic absorption with E(B-V) of 0.07
\citep{vandenbergh1991}. In that case, we find the age of the more
recent starburst drops to 40 Myrs, while the older populations is
characterized by an age of 1 Gyr, and $\chi^{2}_{\nu}$ of 9.0. The
M33 reddening, parameterized in terms of a starburst absorption
curve, remains small, but is actually somewhat higher E(B-V)=0.11
than in the case in which Galactic reddening is ignored. As shown
in figure \ref{fig_nuc_2gal}, there are no large departures
between the model and the data that cannot be directly attributed
either to the interstellar lines, e.g. Mg II, which are not part
of the model, or the coarseness of resolution of the model
spectra.  In these final fits, about 10\% of the mass of the
nucleus would have been involved in the recent starburst. The
masses required in the 10 Myr and 1.1 Gyr starbursts are about
9000 \MSOL\ and 76,000 \MSOL.

To summarize, our analysis implies that there are at least two
stellar populations in the nucleus and that the younger of these
has an age of 40-50 Myrs.  The younger population completely
dominates the FUV light from the nucleus of M33 and contributes
almost half the visible light from the (central part of the)
nucleus. This analysis therefore generally confirms those of
\cite{oconnell1983} and \cite{schmidt1990} based on optical
spectra and limited UV data.

There is no need, based on the HST spectra, to invoke a large
amount of internal reddening in the nucleus, as suggested by
\cite{gordon1999}. This is not quite the same, however, as proving
that there is no M33 reddening, since the our best-fit model using
a starburst absorption law suggests that only 38\% and 66\% of the
light emitted at 2000 and 5700) \AA\ emerges from M33. If the
absorption is between the nucleus of M33 and ourselves, the
implied optical depth is less than 1 at 2000 \AA\ and 0.4 at 5700
\AA. But if the the absorption is within the nucleus itself then
the optical depth through the entire nucleus can approach 1 at
visible wavelengths.


The stellar population history of the region surrounding the
nucleus of M33 is complex, as evidenced by the studies of
\cite{minniti1993} and \cite{mighell1995}.  There are OB
associations with ages of about 5 Myrs, a group of luminous AGB
stars  formed of order 1 Gyrs ago, as well as a considerably older
($>$10 Gyr) metal poor population. \cite{schmidt1990} argued for
three similar populations in the nucleus, one aged less than 5
Myrs, a second 1-5 Gyrs, and a third $>$10 Gyrs, and so it is
possible, perhaps likely, that there is a close connection between
the history of the the nucleus and the nuclear region.  But
definitive proof of that assertion will likely require a more
sophisticated modelling of the HST spectra than in our first
reconnaissance, with stellar models constructed at the resolution
of the HST data and probably the inclusion of comparable quality
data of the nucleus at longer wavelengths.

\subsection{Classification of the Blue Star}

The nature of the star NNW of the nucleus is important for our
purposes because the astrometric accuracy of all X-ray telescopes,
with the possible exception of Chandra, is not sufficient to
distinguish a nuclear source and a source 1 arcsec away.  Indeed
most if not all X-ray imaging surveys of M33 have used the nuclear
X-ray source to register the other sources in the field.

In an attempt to establish the spectral type of the star near the
nucleus of M33, we first compared the UV spectrum to those found
in IUE spectral atlases of O and B stars
\citep{walborn1985,walborn1995}.  These visual comparisons suggest
a spectral classification between O8 and B0.  Stars of spectral
type earlier than O8 show much more prominent \ion{N}{5} and
\ion{Si}{4} profiles that are very different in strength from
those observed in the star in M33. Stars later than B0 have weaker
line profiles than is observed in the M33 star. Assuming a
spectral type of B0, a luminosity class between II and III was
indicated, based on the IUE spectra of HD43818 and HD48434.  Main
sequence stars have lines that are generally narrower than those
observed in our star. Similar results are obtained from a
comparison to the atlas used by \cite{leitherer1999} for spectral
synthesis.  As indicated in Figure~\ref{Star.RoLeHe_o9ii}, which
is a comparison between the reference spectral template for an O9
III, the UV spectrum of the star is not unusual for a late type O
giant.

\cite{massey1996} used the Ultraviolet Imaging Telescope (UIT) on
the Astro-2 space shuttle mission along with ground-based and HST
observations to construct a catalog of 356 sources in M33 that is
thought to be complete to F$_{\lambda}$ of \EXPU{2.5}{-15}{\OIGS},
about 3 times that of the star in M33.  (Given that our star is
about 3 times fainter than the completeness limit and that the
angular resolution of UIT was several arcsec, it is not surprising
that our star is not in that catalog.) There are no late O or
early B giants in this survey; the 5 O9-B0 supergiants have visual
magnitudes that range from 16.8-18.2, which should be compared to
the visual magnitude of our star of 19.25 from our earlier HST
imaging observations or 19.45 from these STIS observations. The
magnitude difference is roughly that expected for a late O or very
early B giant (II-III). Equivalently, the absolute visual
magnitude (ignoring reddening) of $\sim$5.2, given a distance
modulus of 24.5$\pm$0.2 \citep{vandenbergh1991}, is consistent
with a late O or very early B giant \cite[see, e.g.][]{allen1973}.
Thus we can be reasonably confident of the classification of the
star.

\section{Discussion} \label{discussion}

Our initial impulse to obtain a spectrum of the nucleus was
motivated by a desire to identify the optical counterpart to the
X-ray source in M33.  Neither the spectrum of the nucleus nor, for
that matter, the spectrum of the nearby star show direct
indication of X-ray activity.

A number of suggestions have been made over the years as to the
nature of the X-ray source at the heart of M33. These have
included a very low luminosity active galaxy, collections of
neutron star binaries with high or low mass companions
\citep{markert1983,oconnell1983}, and, more recently, a binary
star system containing an Eddington-limited 10 \MSOL\ black hole
\citep{takano1994}.  This last explanation is clearly the favored
explanation today as more point-like, super-Eddington sources have
been found in and out of the nuclei of normal galaxies \cite[see,
e.g.][]{makishima2000}, since they were originally noted by
\cite{long1983}. The X-ray spectrum of the M33 source is similar
to these other sources.

\subsection{A mini-AGN?}

Despite recent suggestions of a jet emerging from the nucleus of
M33 \citep{matthews1999}\footnote{Our inspection of the HST images
of the ``jet'', which was identified in V and R, indicates that it
is almost certainly a small number of stars, one of which is the
blue star whose spectrum we discuss here.}, the existence of
mini-AGN such as NGC4395 \cite[a Sd galaxy with a Seyfert nucleus
with an $L_{X}$ of $10^{38} ~ $ergs~s$^{-1}$;][]{lira1999}, and
clear evidence of a supermassive black hole (SMBH) at the center
of our Galaxy , there is little to support a mini-quasar
interpretation of M33 X-8:

(1) There is no indication whatsoever of AGN-like activity in our
FUV spectra. There is no evidence of broad emission lines, nor for
any blue bump which one might associate with the accretion disk.
Based on the observed X-ray flux
(\EXPU{7.6}{-12}{erg~cm^{-2}~s^{-1}~keV^{-1}} in the 0.1-2.4~keV
band; Long et al 1996), we would expect a far-UV flux of
\EXPU{\sim1-6}{-15}{erg~cm^{-2}~s^{-1}~\AA^{-1}} for an assumed
X-ray/optical spectral slopes $\alpha_{ox}$ of 0.65--1
\citep{wilkes1987}. This would represent a  significant fraction
of the observed flux and yet the UV spectrum is entirely
consistent with that of normal early-type stars. Values of
$\alpha_{ox}$ in this range would be expected from a direct
extrapolation of the X-ray-derived spectral index on quasars
\citep{wilkes1987}. But in fact, such an extrapolation is
conservative; the observed X-ray/optical spectral slopes for
quasars and Seyfert 1 galaxies are more typically 1.5
\citep{mushotzky1989}, implying a larger amount of
quasar-associated UV light. Lower luminosity AGN, including
NGC4395, tend, if anything, toward slightly higher spectral
indices \citep{lira1999}. The amount of absorption along the line
of sight to M33 X-8  based on our analysis of the UV/optical
spectra of the nucleus and X-ray estimates of the column density
of N$_H$ of $\sim10^{21} \: cm^{-2}$ to M33 X-8 itself
\cite[e.g.][]{long1996}, and so emission from a mini-AGN would
seem to be difficult to hide.

(2) Furthermore the mass contained within the M33 nucleus is
simply too small to power a mini-AGN. Even a decade ago
\cite{kormendy1993} had limited the mass of any central SMBH to
\EXPU{<5}{4}{\MSOL} from the velocity dispersion of the nucleus.
More recently, \cite{lauer1998} reduced this limit to
\EXPU{<2}{4}{\MSOL} with higher spatial resolution HST imagery,
and most recently \cite{gebhardt2001} lowered it further to
$<1500~\MSOL$! Hence the M33 nucleus is totally unlike the SMBH in
the center of our Galaxy.



\subsection{Or Black-Hole X-ray Binary/ULX?}

Therefore, we believe that a single X-ray binary  must provide the
explanation for X-8.  It cannot be multiple because, using data
obtained over 5 years with ROSAT, we discovered a 20\% modulation
of the nuclear X-ray flux with a period of 106 d
\citep{dubus1997}. Assuming the L$_X$ of
$\sim$10$^{39}$erg~s$^{-1}$ is produced by an Eddington-limited
source, a compact object mass of $\sim10\MSOL$ is required. This
is in the middle of the range of dynamically-determined masses for
the black-hole X-ray binaries \cite[see, e.g.][]{charles2001}. If
this is correct, then M33 X-8 is analogous to the large number of
comparably bright (but off-nuclear) sources now being found by
Chandra and XMM within nearby spirals \cite[see,
e.g.][]{king2001}, rather than an AGN nucleus.

However, the great majority of the BHXRBs have low-mass stellar
donors (and hence are LMXBs) and are X-ray transient sources
\cite[see e.g.][]{king2000}, a behavior very different from the
remarkably steady (on a $\sim$20 year timescale) X-ray output of
M33 X-8.  This suggests that the companion may be massive, in
which case the source is more analogous to Cyg X-1, which is of
comparable mass and not transient. Cyg X-1 also has a
``superorbital'' period of 142$\pm$7 d, which is interpreted as
precession of the disk in the system \citep{brocksopp1999}. As a
result, we suggested that the BH companion is a giant orbiting
with a period of $\sim$10 days and the 106 d modulation is
``superorbital"  \citep{dubus1997}.

A HMXB might also be expected given the existence of a cluster of
early-type stars in the M33 nucleus resulting from recent
starburst activity.  There would have been about 200 stars more
massive than 8 $\MSOL$ created in the 40 Myr starburst that exists
in M33.  If this picture is correct, then the location of M33 X-8
in the nucleus may be fortuitous, and of more importance is the
association with recent star formation.

In this case, the history of the binary system forming M33 X-8 is
more akin to the ultra-luminous X-ray sources (ULX) identified
with young associations in e.g. the Antennae \citep{fabbiano2001}.
However, M33 X-8 is one of the lower luminosity ULX sources, since
many of them approach $\sim$10$^{40}$erg~s$^{-1}$, for which
compact object masses of $\sim$100 \MSOL\ are being inferred
\cite[e.g.][]{makishima2000}. There are several points to be made
about these mass estimates, which are based upon an
Eddington-limiting luminosity for a uniformly radiating source:

(1) As noted by \cite{makishima2000} and \cite{watarai2001},
although multi-color disk models provide good fits to X-ray
spectra of ULX sources, the disk temperatures are in the range
1.1--1.8~keV, which is much too high for $\sim$100\MSOL\ BHs.
\cite{king2001} have proposed a specific model involving mild
beaming (essentially the opening cone angle of the inner accretion
disc) to reduce the implied compact object mass.

(2) There are also clearly some sources that exceed the Eddington
limit. For example, we already know that neutron star XRBs, such
as A0538-66, the periodically recurrent transient in the LMC, are
capable of reaching $\sim$10$^{39}$~erg~s$^{-1}$ \citep[see
e.g.][]{charles1983}.  And since this is a fast X-ray pulsar
($\sim$70 ms) this is presumably associated with the highly
non-spherical magnetic accretion of the neutron star geometry. (We
note in passing that there are several other XRBs in the LMC that
are more luminous than their galactic counterparts. For example
\cite{clark1978} have suggested that this is associated with the
lower metallicity of the LMC.)

The micro-quasar X-ray transient GRS~1915+105 may be the closest
analogue to M33 X-8 in our Galaxy. While GRS~1915+105 is labelled
as a ``transient'' X-ray source, it has actually been ``on'' now
for a decade, and we believe that the few decades lifetime of
X-ray astronomy is much too short to make clear-cut distinctions
in the ``steady/transient'' label for many objects.

Early high mass estimates for the BH in GRS~1915+105 now appear to
be incorrect. \cite{greiner1998} showed with RXTE X-ray
spectroscopy that this source with a peak X-ray luminosity of
$\sim$7$\times$10$^{39}$~erg~s$^{-1}$ in the 1-200~keV band (or
even greater if the lower limit for the energy band was taken down
to 0.1~keV) fully qualifies as an ULX with an implied
luminosity-estimated mass of order 70 \MSOL.
\cite{gliozzi1999} pointed out that the minimum bulk kinetic power
of $>$3$\times$10$^{40}$~erg~s$^{-1}$ during the jet ejection
events far exceeded the observed peak X-ray luminosity, and hence
inferred a very massive compact object as the energy source.
Furthermore the observation of 67 Hz X-ray QPOs in GRS~1915+105
was interpreted as evidence for a $\sim$33\MSOL\ compact object
\citep{morgan1997}. However, following their identification
\citep{greiner2001} of the spectral type of the mass donor in GRS
1915+105, \cite{greiner2001b} announced the discovery of the
orbital period (33.5 d) and radial velocity curve of the IR
counterpart to the X-ray source. This implies a compact object
mass of 14$\pm$4~\MSOL, very similar to, but at the upper end of,
the masses of the other BHXRBs.  It should be noted that Greiner's
IR spectrum implies a low-mass ($\sim1.2$ \MSOL) but evolved (K-M
giant) companion, not the early-type companion that had been
suggested in previous studies.

Consequently, we suggest that the properties of M33 X-8 and
GRS~1915+105 can be used to draw useful comparisons with the ULX
population. Based on the evidence at hand, all may be explainable
as $\sim$10--20\MSOL\ BHs.

\subsection{But is it possible that the O star near the nucleus is
the optical counterpart?}

Our bias is to associate the M33 X-8 with the nucleus of M33,
rather than the O star located NNW of the nucleus. Our bias, as
has been the bias of most if not all of those interested in M33
X-8, is based upon its positional coincidence with the nucleus and
the fact it is so luminous compared to other sources in M33 (and
the local group).  The spectra we have obtained of the O star have
not provided evidence that undercut our prejudices on this issue.
There are a large number of late O stars spread over M33; there
remains only one nucleus.

If the star were the optical counterpart, the X-ray to optical
luminosity ratio of $\sim$30 would be extreme for an X-ray system
involving a high mass companion and therefore one might expect
that the effect of the X-ray source on the companion would also be
extreme.  But most optical counterparts are revealed by time
variations in either the broad band continuum due to heating of
the companion's photosphere or to phase-dependent variations in
line profiles produced by ionization of the companion's wind
\cite[see, e.g.][]{vanloon2001}. As we noted, the spectrum of the
star NNW of the nucleus appears normal for a late type O giant.
But our observations were not structured for searches for orbital
phase dependent effects, given typical orbital period of massive
X-ray binaries, and therefore it is not possible to completely
rule out the O star as the optical counterpart.

The question of where M33 X-8 is located could and should be
solved using Chandra, although not with the existing CCD
observations, which are saturated at the position of the nucleus
(Jonathan McDowell 2001, private communication).

\section{Conclusions} \label{conclusions}

In summary, we have carried out the first high S/N UV and FUV
spectroscopic study of the nucleus of M33.  Our analysis confirms
that at least two populations of stars exist in the nucleus, and
that star formation occurred in the nucleus 40-50 Myrs ago. The
X-ray source in the nucleus of M33 does not affect the UV spectrum
significantly, all of which is consistent with the absence of a
supermassive black hole or mini-quasar at the center of the
nucleus, and the hypothesis that the brightest X-ray source in the
Local Group is a BH with a high-mass companion.  M33 X-8 will
remain interesting as the nearest ULX, even though it is hidden at
UV and visible wavelengths in the light of the relatively
unreddened nucleus of M33.

\acknowledgments{This work was supported by NASA through grant
G0-6544 from the Space Telescope Science Institute, which is
operated by AURA, Inc., under NASA contract NAS5-26555. GD
acknowledges additional support from NASA grants NAG5-7007 and
NAG5-7034. Brian Monroe, a Data Analyst at STScI, helped us to
reduce the data for this project. }


\pagebreak

\figcaption[m33_nuc_data.ps] {The flux-calibrated spectrum of the
nucleus of M33. The more prominent absorption lines are labelled
and the positions of the Balmer lines are indicated.
\label{fig_nuc_data}}

\figcaption[m33_star_data.ps] {The flux-calibrated spectrum of the
star NNW  of the nucleus M33. The stronger stellar and
interstellar features are labelled.  Note that the vertical scales
differ in the three panels. \label{fig_star_data}}



\figcaption[nuc_fits_single_bw.01nov.ps] {A comparison between the
spectrum of the nucleus and the ``best-fitting'' single component
models. In the upper panel, the best-fit assuming a starburst
absorption law is offset above the data.  The best fit assuming a
Galactic absorption law is plotted below the data. The lower panel
shows the difference between the best-fitting single component
model with a starburst absorption law and the data.
\label{fig_nuc_1}}

\figcaption[double_bw.01nov26.ps] {A comparison between the
spectrum of the nucleus and the ``best-fitting'' two component
normal abundance synthesis model.  The higher resolution UV
spectra were interpolated into the model spectra in the fits shown
here.  A starburst reddening curve was assumed for the M33
nucleus; the associated E(B-V) of the best-fit was 0.06.  The
contribution of the 50 Myr and 1 Gyr bursts of star formation are
also shown in the upper panel. The difference spectrum
(model-data), shown in the bottom panel shows evidence of a  2200
\AA\ feature. \label{fig_nuc_2}}

\figcaption[double_gal_bw.01nov26.ps]{A comparison between the
spectrum of the nucleus and the ``best-fitting'' two component
normal abundance synthesis model, when allowing for a fixed
Galactic extinction of E(B-V)=0.07 and separate M33 absorption
parameterized in terms of a starburst reddening curve. The
contributions of the 40 Myr and 1 Gyr starbursts are also shown in
the upper panel. The fitted reddening associated with M33
corresponds to an E(B-V)=0.11.  The difference spectrum in the
lower panel indicates there are no significant departures between
the data and the model including Galactic absorption that cannot
be accounted for by the limited resolution of the model spectra or
interstellar lines. \label{fig_nuc_2gal} }

\figcaption[Star.RoLeHe_o9ii.ps] {A comparison between the UV
spectrum star NNW  of the nucleus M33 compared to the O9 II
template from the atlas used by
\cite{leitherer1999}.\label{Star.RoLeHe_o9ii}}

\clearpage
\begin{center}
\begin{deluxetable}{lcccr}
\tablecaption{Journal of HST STIS Observations of M33 Nucleus \label{tab_obs}}
\tablehead{\colhead{Observation~Date} &
 \colhead{Detector} &
 \colhead{Grating} &
 \colhead{Wavelength~Range} &
 \colhead{Exposure}
\\
\colhead{~} &
 \colhead{~} &
 \colhead{~} &
 \colhead{(\AA)} &
 \colhead{(s)}
}
\scriptsize
\tablewidth{0pt}\startdata
1999-Oct-06 &  FUV-MAMA &  G140L &  1150-1736 &  10,157 \\
2000-Oct-20 &  NUV-MAMA &  G230L &  1570-3180 &  2,810 \\
2000-Oct-20 &  CCD &  G430L &  2900-5700 &  2,128 \\
\enddata
\end{deluxetable}
\end{center}

\newpage
\pagestyle{empty}
\begin{figure}
\plotone{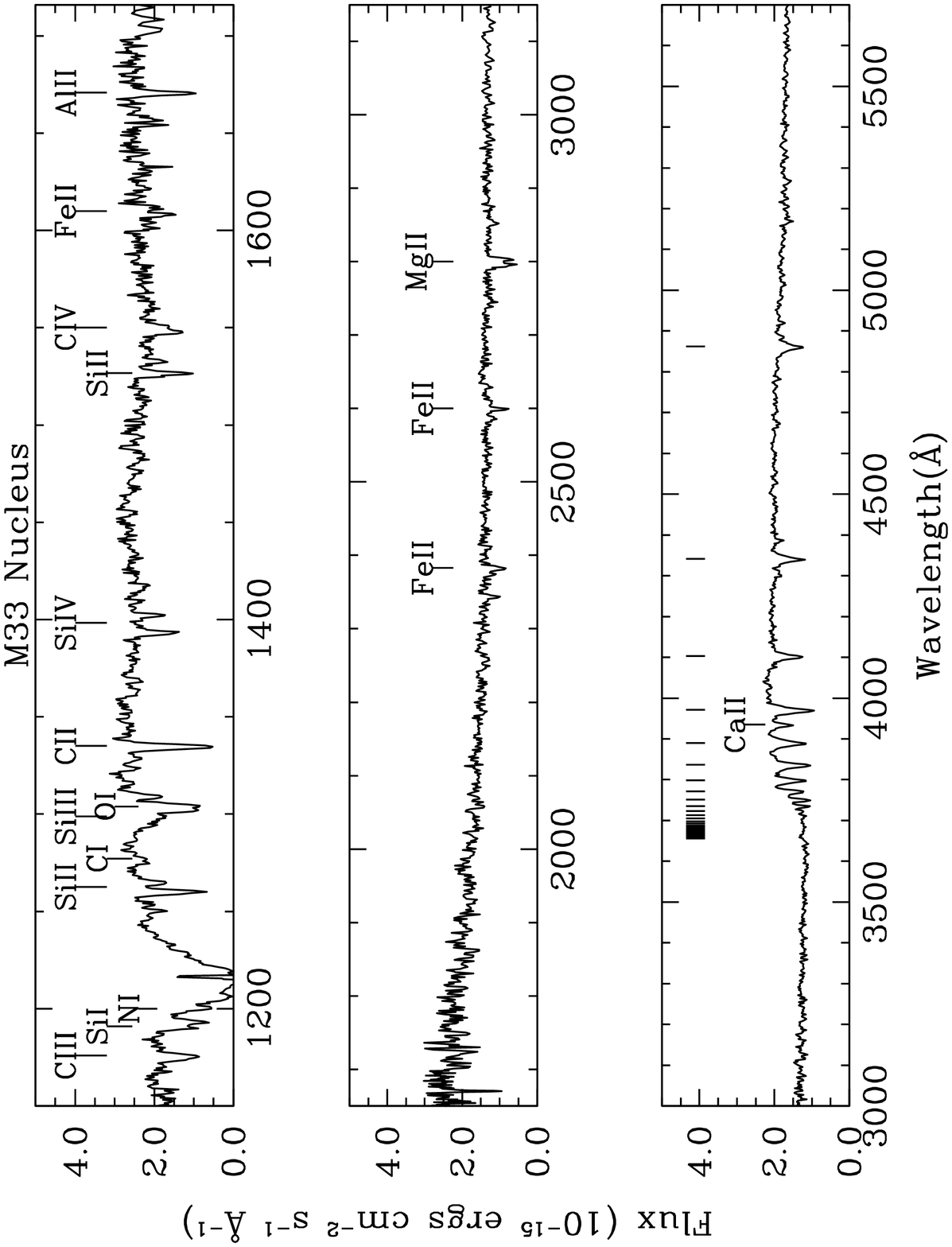}
\end{figure}

\newpage
\pagestyle{empty}
\begin{figure}
\plotone{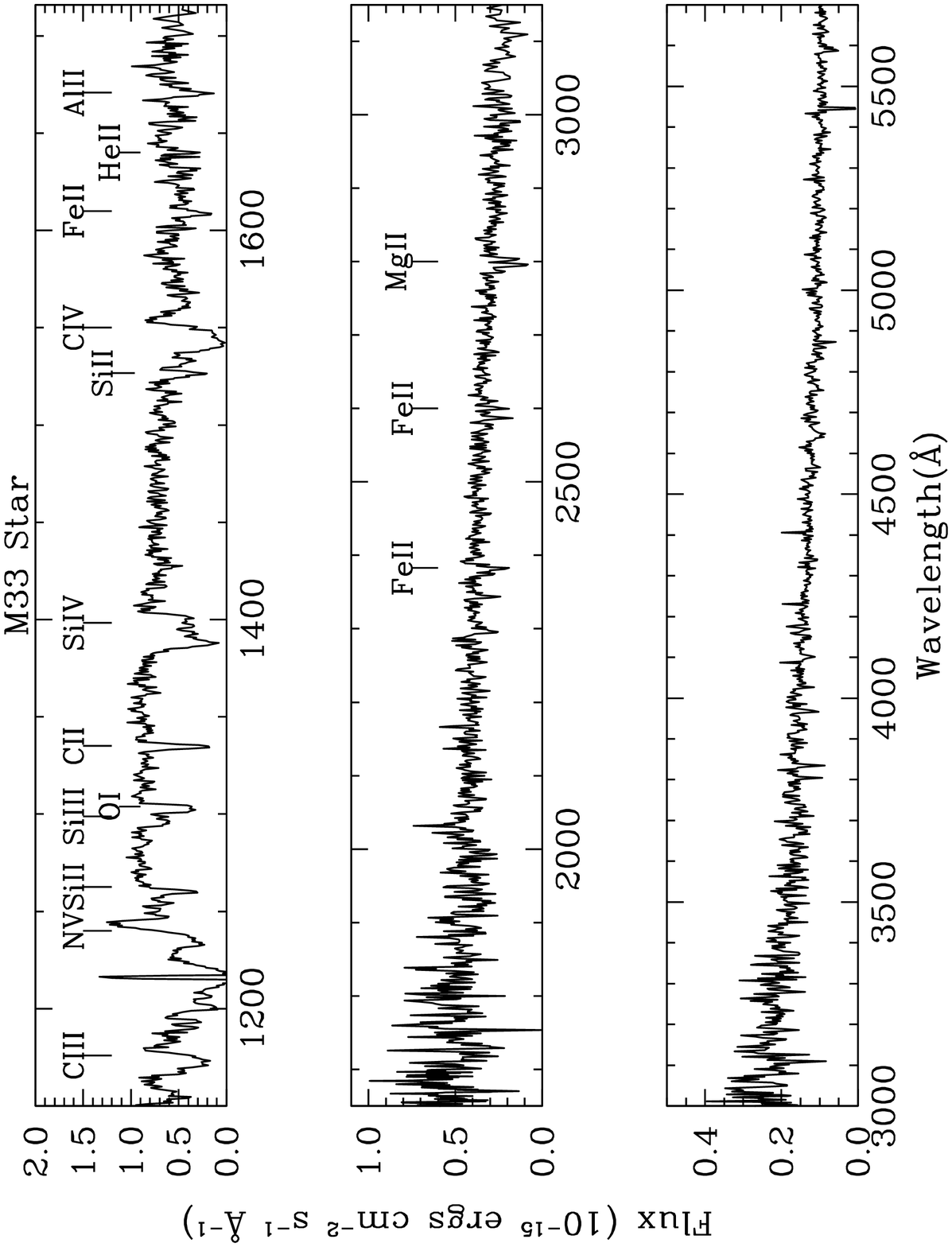}
\end{figure}

\newpage
\pagestyle{empty}
\begin{figure}
\plotone{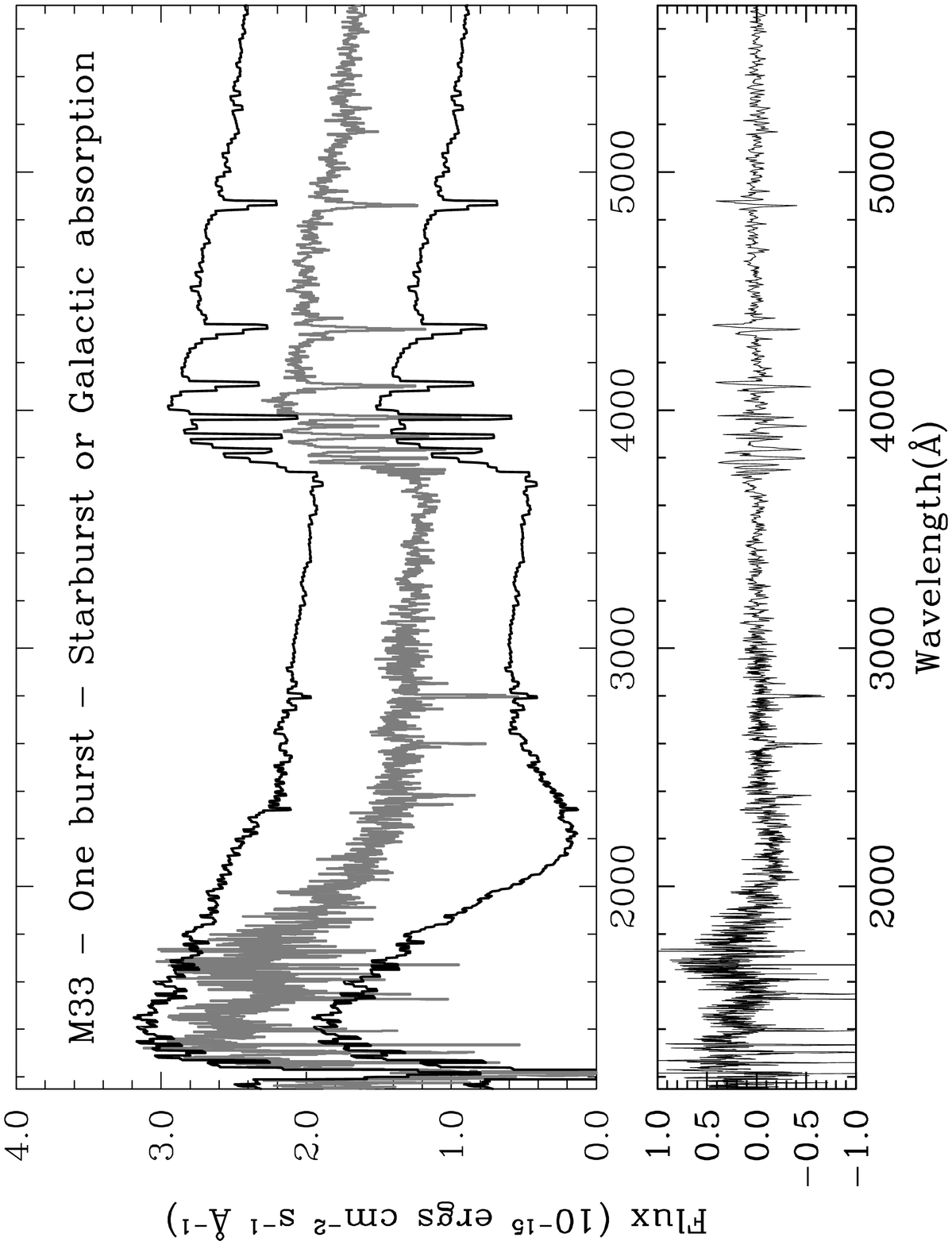}
\end{figure}

\newpage
\pagestyle{empty}
\begin{figure}
\plotone{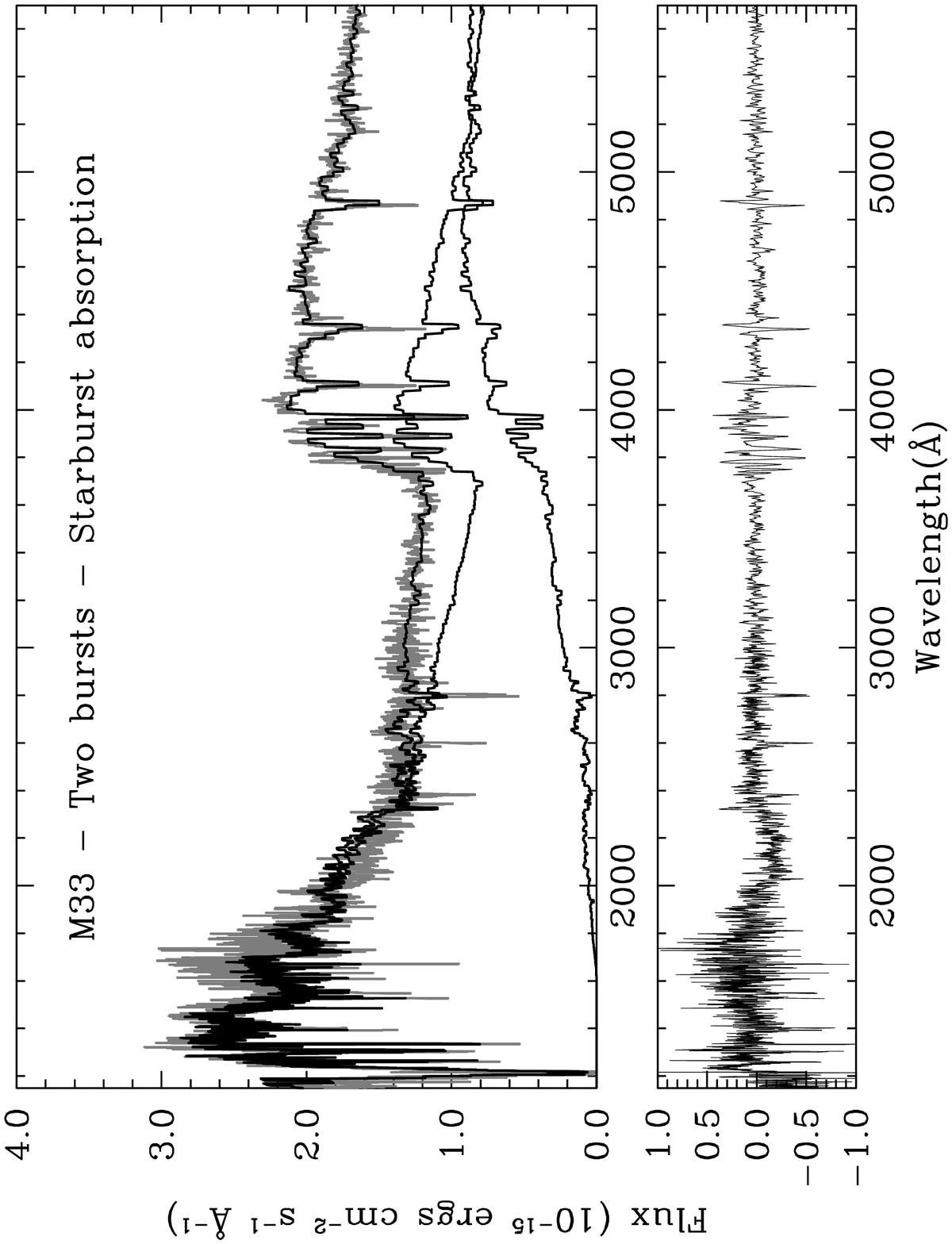}
\end{figure}

\newpage
\pagestyle{empty}
\begin{figure}
\plotone{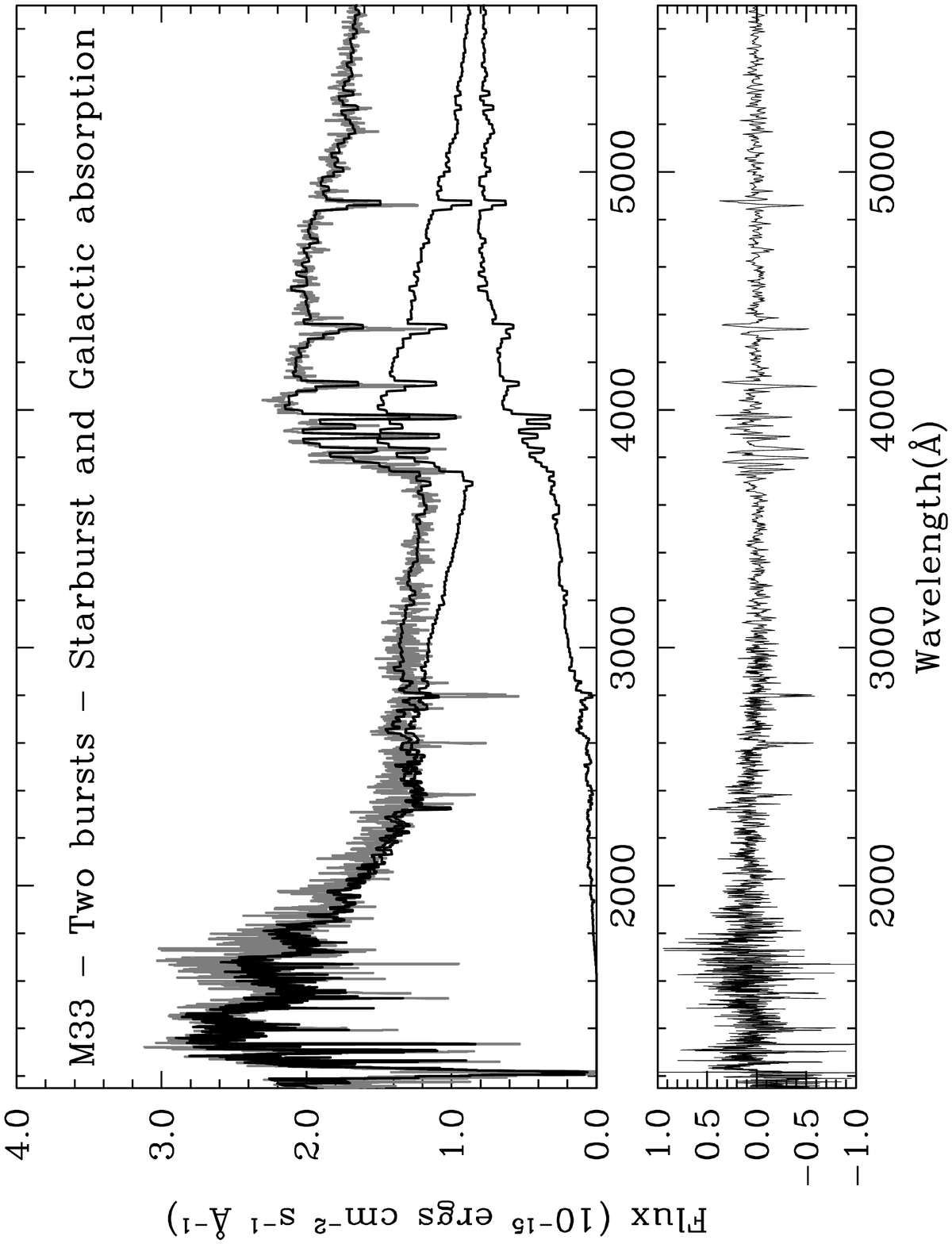}
\end{figure}

\newpage
\pagestyle{empty}
\begin{figure}
\plotone{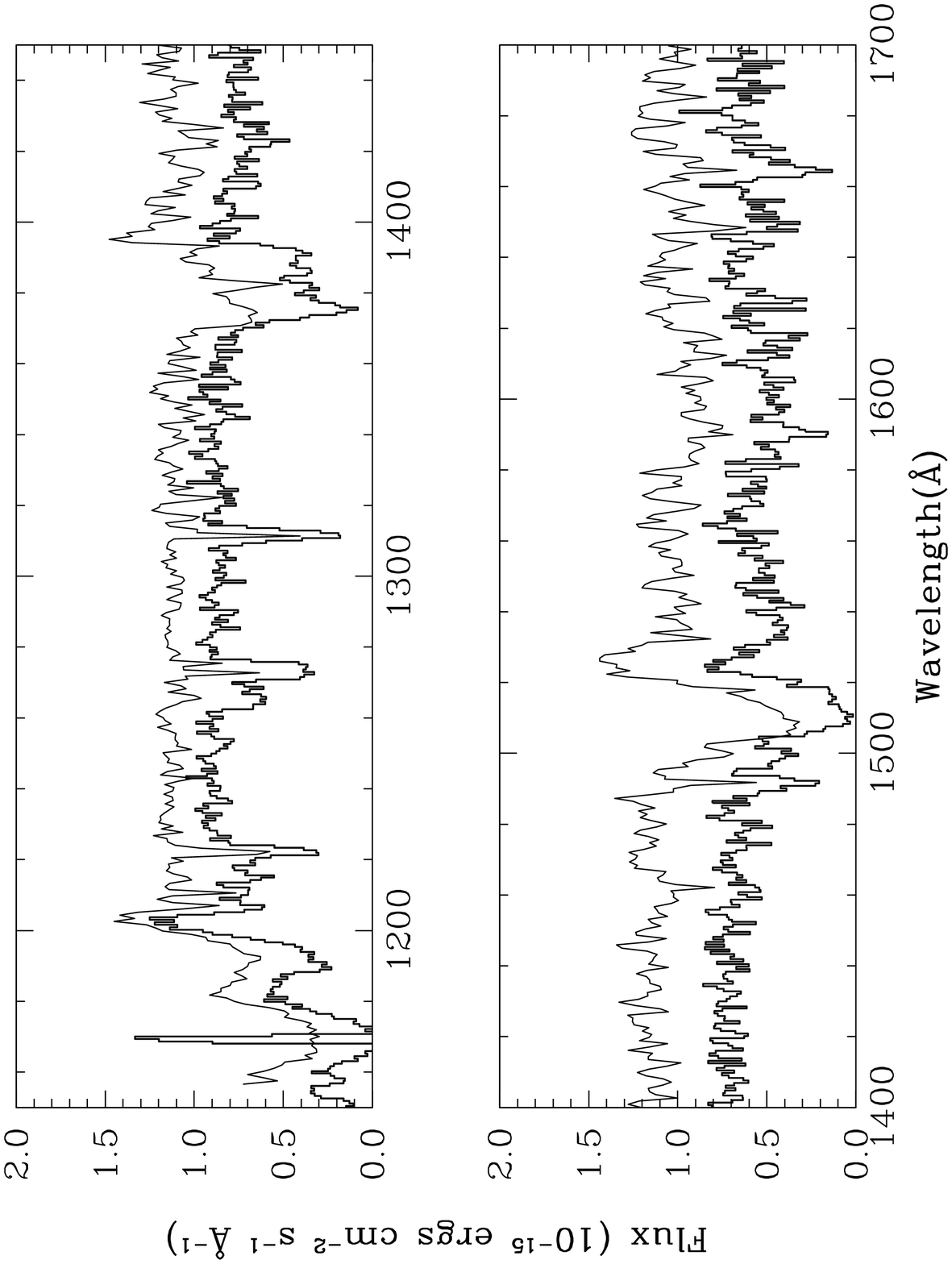}
\end{figure}


\end{document}